\documentclass[sigplan,screen,authorversion,nonacm]{acmart}
\usepackage{microtype}
\usepackage{graphicx}
\usepackage{subcaption}
\usepackage{booktabs}
\usepackage{hyperref}
\usepackage{tikz}
\usepackage{xcolor}
\usepackage{multirow}
\usepackage{amsmath}
\usepackage{listings}

\usepackage{array}
\newcolumntype{R}[1]{>{\raggedleft\let\newline\\\arraybackslash\hspace{0pt}}m{#1}}

\DeclareRobustCommand{\colordot}[2][red,fill=red]{\tikz[baseline=-.7ex]\draw[#1,radius=#2] (0,0) circle ;}%
\definecolor{magmapurple}{RGB}{70,3,159}
\definecolor{magmaorange}{RGB}{251,159,58}

\AtBeginDocument{%
  }

\copyrightyear{2023}
\acmYear{2018}
\acmDOI{XXXXXXX.XXXXXXX}
\setcopyright{none}

\begin{document}

\title[Performance Embeddings]{Performance Embeddings: A Similarity-based Approach to Automatic Performance Optimization}

\author{Lukas Tr\"umper}
\email{lukashans.truemper@inf.ethz.ch}
\affiliation{%
  \institution{ETH Zurich}
  \country{Switzerland}
}

\author{Tal Ben-Nun}
\authornote{Work on this paper was done while at ETH Zurich.}
\email{talbn@llnl.gov}
\affiliation{%
  \institution{Lawrence Livermore National Laboratory}
  \country{USA}
}

\author{Philipp Schaad}
\email{philipp.schaad@inf.ethz.ch}
\affiliation{%
  \institution{ETH Zurich}
  \country{Switzerland}
}

\author{Alexandru Calotoiu}
\email{alexandru.calotoiu@inf.ethz.ch}
\affiliation{%
  \institution{ETH Zurich}
  \country{Switzerland}
}

\author{Torsten Hoefler}
\email{htor@inf.ethz.ch}
\affiliation{%
  \institution{ETH Zurich}
  \country{Switzerland}
}

\renewcommand{\shortauthors}{Tr\"umper et al.}

\begin{abstract}
Performance optimization is an increasingly challenging but often repetitive task. While each platform has its quirks, the underlying code transformations rely on data movement and computational characteristics that recur across applications. %
This paper proposes to leverage those similarities by constructing an embedding space for subprograms. The continuous space captures both static and dynamic properties of loop nests via symbolic code analysis and performance profiling, respectively.
Performance embeddings enable direct knowledge transfer of performance tuning between applications, which can result from autotuning or tailored improvements. 
We demonstrate this \textit{transfer tuning} approach on case studies in deep neural networks, dense and sparse linear algebra compositions, and numerical weather prediction stencils.
Transfer tuning reduces the search complexity by up to four orders of magnitude and outperforms the MKL library in sparse-dense matrix multiplication.
The results exhibit clear correspondences between program characteristics and optimizations, outperforming prior specialized state-of-the-art approaches and generalizing beyond their capabilities.
\end{abstract}


\maketitle

\begin{figure}
    \centering
    \includegraphics[width=\linewidth]{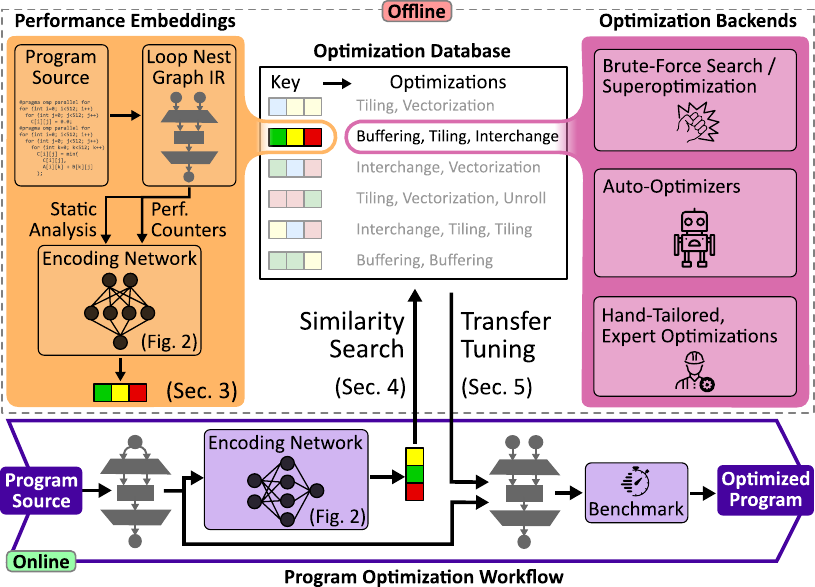}
    \vspace{-1.7em}
    \caption{An overview of the similarity-based approach to automatic performance optimization.}
    \label{fig:Overview}
    \vspace{-0.8em}
\end{figure}

\section{Introduction}
\label{sec:Introduction}

Automatic performance optimization of programs for modern computing architectures is challenging.
Even for smaller programs, the possibilities to schedule the operations and the data movement become infeasible to exhaustively explore. 
To efficiently navigate the optimization space, a performance model could be constructed as a surrogate to approximate the search; the searched parameters can be limited to a small number for brute-force tuning; or, more often than not, the program is optimized manually by a performance engineer.

Several performance models have been developed for specific program classes, notably Polyhedral subprograms~\cite{polyhedral}.
The polyhedral model has helped in developing several automated tuning methods based on integer-linear programming~\cite{Baghdadi:2013} and machine learning~\cite{Chen:2018, Adams:2019, Baghdadi:2021} as well.
Such methods primarily target optimizations on the loop level such as interchanging their order and tiling the iteration space.
However, due to the need for expressing programs with affine array accesses and simple loop bounds, these techniques are limited in representing real-world applications.

Methods used for optimizing data-dependent applications, such as sparse linear algebra routines, have to rely on specialized, input-specific models~\cite{Elafrou:2017}.
Because such models are hard to integrate into a general tuning framework, performance engineers often fall back to general profiling-based performance models, such as the roofline model~\cite{Williams:2009}, for custom applications.
Since profiling-based models lack a connection to the algorithmic structure, their interpretation requires significant experience~\cite{Treibig:2014}, which makes the search for optimizations hard to automate.
Optimization efforts for real-world applications are thus often resource-intensive manual processes where the outcome strongly depends on the skill set of the individual performance engineer~\cite{Carlson:2020, Takahashi:2021, BenNun:2022}.

In this paper, we present a similarity-based approach to the automatic performance optimization of general loop nests, summarized in Figure~\ref{fig:Overview}.
We develop a method for encoding both static and dynamic performance characteristics of loop nests and capturing them as \emph{performance embeddings} --- a latent, continuous space in which a multidimensional point represents a subprogram.
Based on these embeddings, which are trained separately, optimizations derived from a variety of methods (such as brute force, manual tuning, or state-of-the-art auto-schedulers) are stored in an optimization database. This enables knowledge transfer of optimization between different programs with similar static or runtime characteristics, which we call \textit{transfer tuning}.

During transfer tuning, loop nests are then optimized by fuzzy matching the optimizations of the k-nearest neighbors from the database according to their performance embeddings.
We demonstrate the effectiveness of our approach on a series of polyhedral and non-polyhedral real-world applications, significantly reducing the search complexity for performance optimizations and outperforming state-of-the-art auto-schedulers by reaching up to \textbf{92\%} better runtime improvements.

In summary, this paper makes the following contributions:
\begin{itemize}
    \item Methodology for encoding performance characteristics of general loop nests in \emph{performance embeddings};
    \item Construction of an extensive optimization database that is separate from the performance modeling;
    \item Reduction of the optimization search space size by orders of magnitude through \emph{transfer tuning};
    \item Demonstration of effectiveness compared with state-of-the-art auto-optimizers and extending the database with tailored optimizations in two case studies.
\end{itemize}

\begin{figure*}
    \centering
    \includegraphics[width=\linewidth]{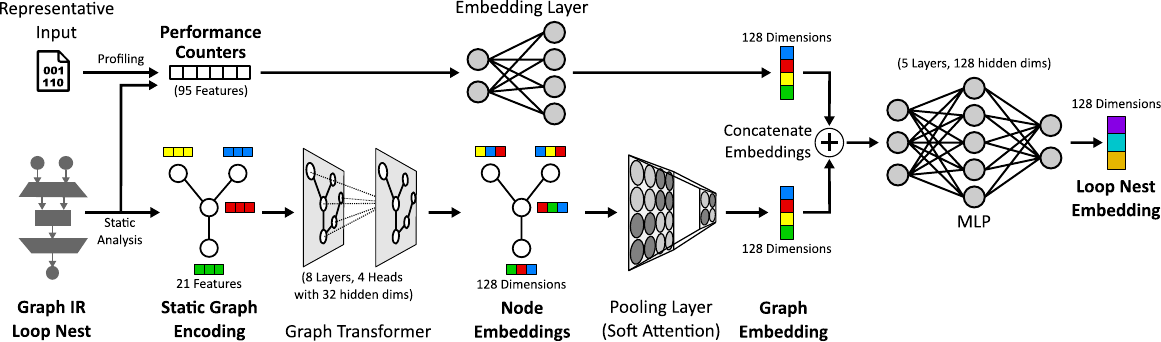}
    \caption{An overview of the model architecture to construct loop nest embeddings.}
    \label{fig:Model_Architecture}
\end{figure*}

\section{Similarity in Performance Optimization}
\label{sec:Background}

Programs with different structural properties may still share similar performance characteristics, which allow them to be optimized in similar manners.

The following example shows a standard matrix multiplication and a min-plus matrix multiplication commonly used for shortest-path problems:
\begin{center}
    \includegraphics[width=\linewidth]{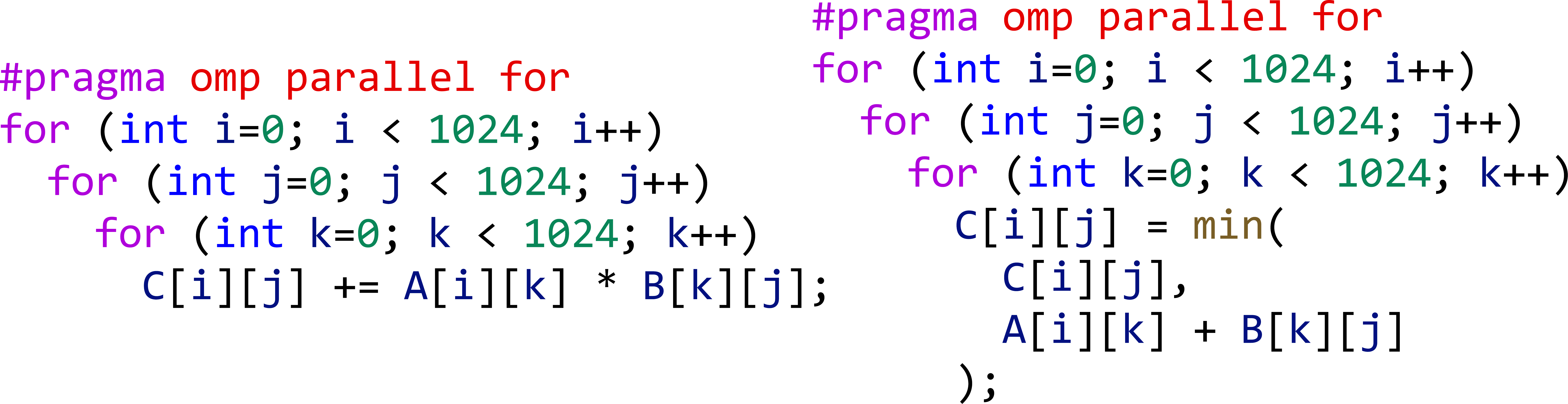}
\end{center}
The loop nests are structurally identical, thus trivially sharing similar performance characteristics.
Consequently, the min-plus matrix multiplication can be optimized using the same tiling, buffering, and vectorization strategies found in the literature for matrix-matrix mutliplication~\cite{Kazushige:2008}.
Due to the structural similarity, existing auto-schedulers based on the polyhedral model are able to detect this reliably~\cite{Adams:2019}, reaching a speedup of up to 26x over the na\"ive version.

The same potential for optimizations can, however, also be observed in a structurally different application, such as the sparse-dense matrix multiplication shown below:
\begin{center}
    \includegraphics[width=.65\linewidth]{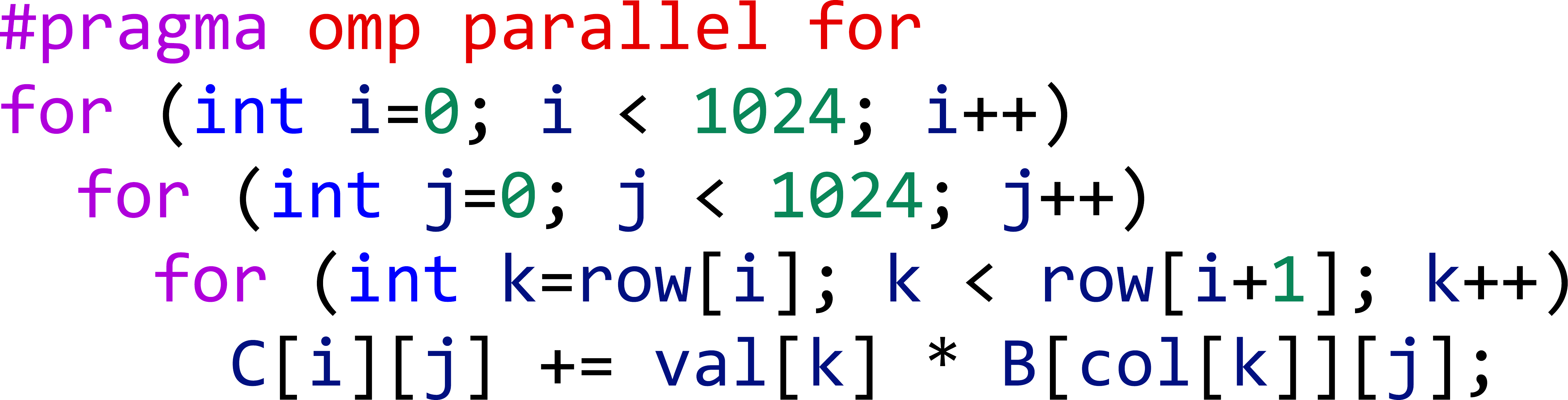}
\end{center}
In comparison to the regular, dense matrix multiplication from before, this loop nest is no longer data-oblivious, since the innermost loop bounds are data-dependent.
However, both programs exhibit a strided memory access to the dense matrix \texttt{B}, which can be resolved by interchanging the two innermost loops to get a speedup of 2.5x.
Existing auto-schedulers~\cite{Adams:2019, Baghdadi:2021} can apply this optimization for the original matrix multiplication, but can only transfer these optimizations to the sparse multiplication if their performance models indicate similar performance characteristics.
This, in turn, is only possible in static models using over-approximation~\cite{Benabderrahmane:2010} or an inspector-executor model~\cite{spf} on the loop bounds for the innermost loop, hindering possible further optimizations with regard to load imbalances.

A case where the structural differences are even more pronounced is shown below, where the first program computes a sparse matrix-vector product, and the second program performs a prime number check on an array of 20,000 numbers:
\begin{center}
    \includegraphics[width=.75\linewidth]{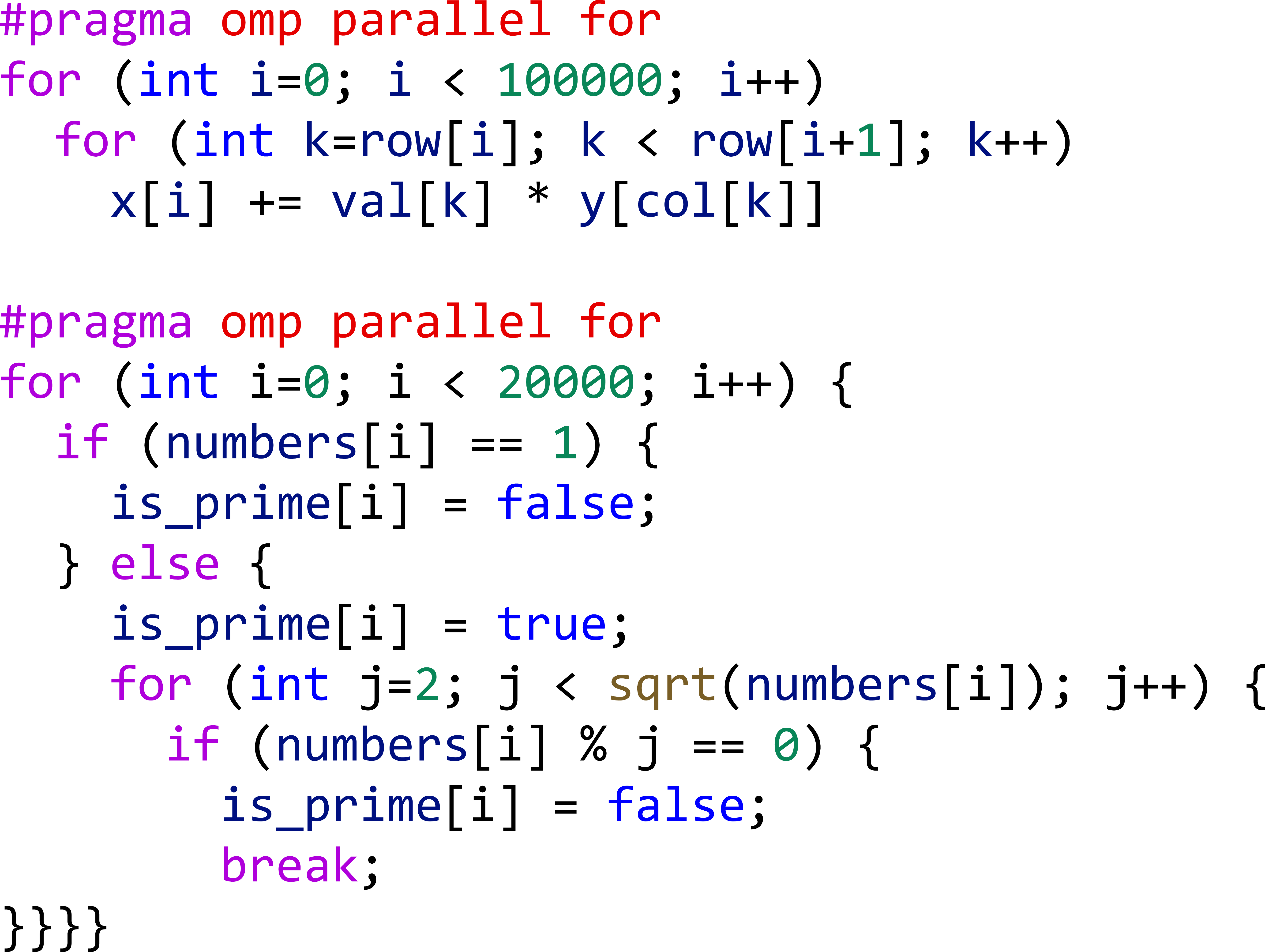}
\end{center}
Despite their structural differences, both programs are inherently prone to an imbalanced distribution of work among different threads when parallelizing the outermost loop.
In both cases, a dynamic assignment of work to threads yields significant speedups of $1.82x$ and $1.49x$ respectively.
While a purpose-built, data-specific model~\cite{Elafrou:2017} can address this problem for the sparse matrix-vector product, the same model cannot directly be applied to the structurally-different prime number filter.
Hence, in order to identify similarities and transfer optimizations between data-dependent applications, the integration of an uncountable number of specialized models is necessary.

In contrast, performance engineers are able to identify similarities between both data-oblivious and data-dependent applications treating data-dependent aspects as \textit{gaps}, which are inferred through profiling.
\emph{Performance embeddings} adopt this observation by encoding both static and dynamic performance characteristics of parallel loop nests, enabling the transfer of optimizations across more general problems.

\section{Embedding Parallel Loop Nests}
\label{sec:Performance_Embeddings}

The basis of the similarity search is a \textit{representation} of parallel loop nests which captures a rich set of performance-relevant properties.
This representation should encode static properties such as the structure of loops, and the data accesses, but also reflect dynamic properties such as the bandwidth utilization, the thread imbalance, or the amount of mispredicted branches.
In contrast to approaches solely focusing on runtime prediction for data-oblivious applications~\cite{Adams:2019, Baghdadi:2021, Singh:2022}, the purpose of this representation is to provide a detailed description of performance for general parallel loop nests; the runtime itself does not expose information about the potential for optimization. 

We compute the representation of parallel loop nests using neural networks based on both static and dynamic features, depicted in Figure~\ref{fig:Model_Architecture}.
Dynamic features (performance counters) measured on representative inputs allow the model to treat input-specific aspects of a parallel loop nest as \textit{gaps} in the static analysis.
These features inform the model about the behavior of the loop nest via hardware metrics.
For example, the load imbalance between threads is a direct result of a matrix's sparsity pattern in a sparse matrix multiplication.

\subsection{Parallel Loop Nests}
\label{subsec:Parallel_Loop_Nests}

Before introducing the representation, the term parallel loop nest shall be defined in detail.
A \textit{parallel loop} defines a parallel iteration space and a (possibly empty) body of computations executed for each iteration.
A \textit{parallel loop nest} is an ordered tree where each node is a parallel loop nested inside the iteration space of the parent.

A program is considered a set of parallel loop nests, which are optimized independently.
This assumes that optimizations on the full program have been determined beforehand, e.g., the identification of parallelism and the fusion of parallel loop nests.
A fusion strategy based on similarity is briefly discussed in Section~\ref{sec:Discussion}.

The computations and loop extents are not assumed to be known at compile-time.
In particular, the body may comprise sequential loops and recursions whose function depends on input data.
Compared with other models~\cite{Adams:2019, Baghdadi:2021}, this definition relaxes the requirements of compile-time known loop extents, operations, and memory access patterns.

\subsection{Encoding}
\label{subsec:Encoding}

The \textit{encoding} maps the parallel loop nest given in \textit{an intermediate representation (IR)} to a set of features, which can be processed by a neural network.
The encoding of parallel loop nests consists of two parts: a graph encoding of the static IR and an encoding of the dynamic profiling information in a single vector.
A detailed list of the used static and dynamic features is presented in Appendix~\ref{sec:Appendix}.

\paragraph{Static Encoding}
The basis of the static encoding is a parallel loop nest represented as a \textit{stateful dataflow multigraph (SDFG)}~\cite{BenNun:2019}.
SDFGs combine state machines with dataflow graphs to represent complete programs, which makes them amenable for static analysis and simplifies the mapping to a graph encoding.
However, the approach could equally be implemented with other IRs, e.g., \textit{LLVM IR}~\cite{Lattner:2004}.

\begin{figure}[t]
    \centering
    \includegraphics[width=.7\linewidth]{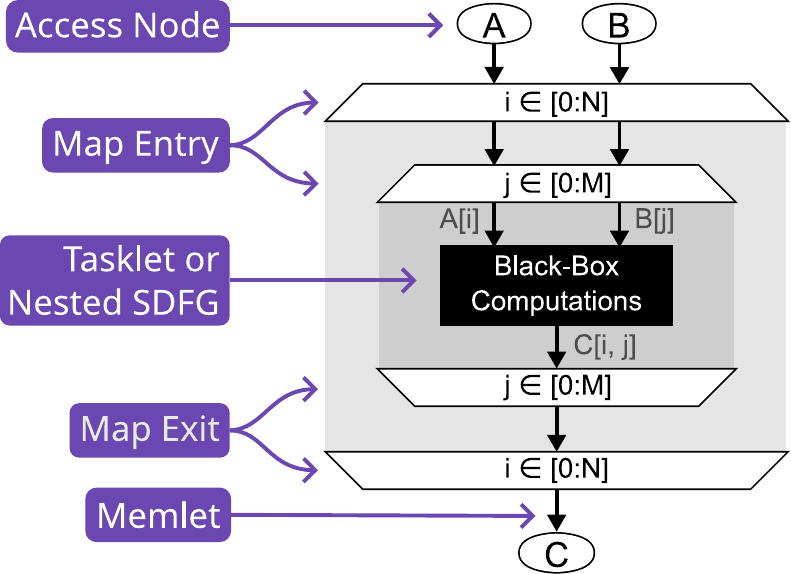}
    \vspace{-1em}
    \caption{SDFG representation of a parallel loop nest.}
    \label{fig:sdfg_syntax}
    \vspace{-.5em}
\end{figure}

At the outermost scope, the SDFG of a parallel loop nest is a dataflow graph comprising at least a single parallel loop, called \textit{map}.
As shown in Figure~\ref{fig:sdfg_syntax}, the body of the map may comprise nested maps, \textit{tasklets} (operations), or nested SDFGs.
The components of an SDFG are mapped to a graph of nodes with features and edges as follows:
\begin{itemize}
    \item \textit{Access node:} Access nodes represent data in the data-flow graph and are mapped to corresponding nodes in the encoding. These nodes are represented by features such as shape, total size, data type, and data layout.
    \item \textit{Map Entry:} A map entry represents the start of the scope of a parallel loop. The map entry is mapped to a node in the encoding and featurized by properties such as the level in the hierarchy, the map extent, and the step size. If the map extent or step size cannot be inferred statically, a special flag is set in the encoding which indicates a dynamic map.  
    \item \textit{Map Exit:} A map exit defines the end of the scope of a parallel loop and is mapped to a node in the encoding represented by one-hot encoding.
    \item \textit{Body node:} The computational nodes inside the body of a parallel loop nest (namely, tasklets and nested SDFGs) are summarized in a single \textit{body node}. The body node is represented by one-hot encoding.
    \item \textit{Memlets:} Memlets are directed edges of the dataflow graph in an SDFG defining the structure of the data accesses. Accordingly, memlets also define the edges of the encoding. In order to collect features for memlets, each edge is split into two edges and an intermediate node encodes the memlet itself. Data accesses are additionally encoded in an \textit{access matrix} following the format of the polyhedral model~\cite{Feautrier:2011}. Non-affine accesses are represented by an empty access matrix and a special flag indicating a non-affine access.
\end{itemize}

\paragraph{Dynamic Encoding}
Processor hardware architectures provide facilities called \textit{performance counters} to collect detailed statistics about the execution of a program.
For example, counters for the total number of executed instructions, or the number of bytes transferred between different levels of the memory hierarchy.
We encode the dynamic profile of a parallel loop nest in a single vector of performance counters for the entire parallel loop nest.
In total, $19$ counters are selected from $8$ different categories: \textit{instructions}, \textit{FP32}, \textit{FP64}, \textit{branching}, \textit{main memory}, \textit{L3 cache}, \textit{L2 cache} and \textit{DRAM controller}.
Each counter is measured for all threads during the profiling and the statistics \textit{min}, \textit{max}, \textit{mean}, \textit{std. deviation}, and \textit{sum} are computed over all threads.
Hence, the resulting vector contains $95$ different features.

\subsection{Model}
\label{subsec:Model}

As illustrated in Figure~\ref{fig:Model_Architecture}, the two encodings are first processed in separate branches of the neural network.
A linear \textit{embedding layer} maps the dynamic encoding to a \textit{dynamic embedding}.
A \textit{graph neural network (GNN)} based on the \textit{graph transformer operator}~\cite{Shi:2021} maps the static encoding to \textit{node embeddings}, which are summarized into a graph embedding by an attentional \textit{pooling layer}~\cite{Li:2015}.
Finally, the graph embedding is concatenated with the dynamic embedding and mapped by another MLP to an embedding of the entire parallel loop nest.
The size of the embeddings is fixed to $128$ for node and graph embeddings.
In total, the model comprises $44$ layers and $862$,$000$ trainable parameters.

\begin{figure}[t]
    \centering
    \includegraphics[width=\linewidth]{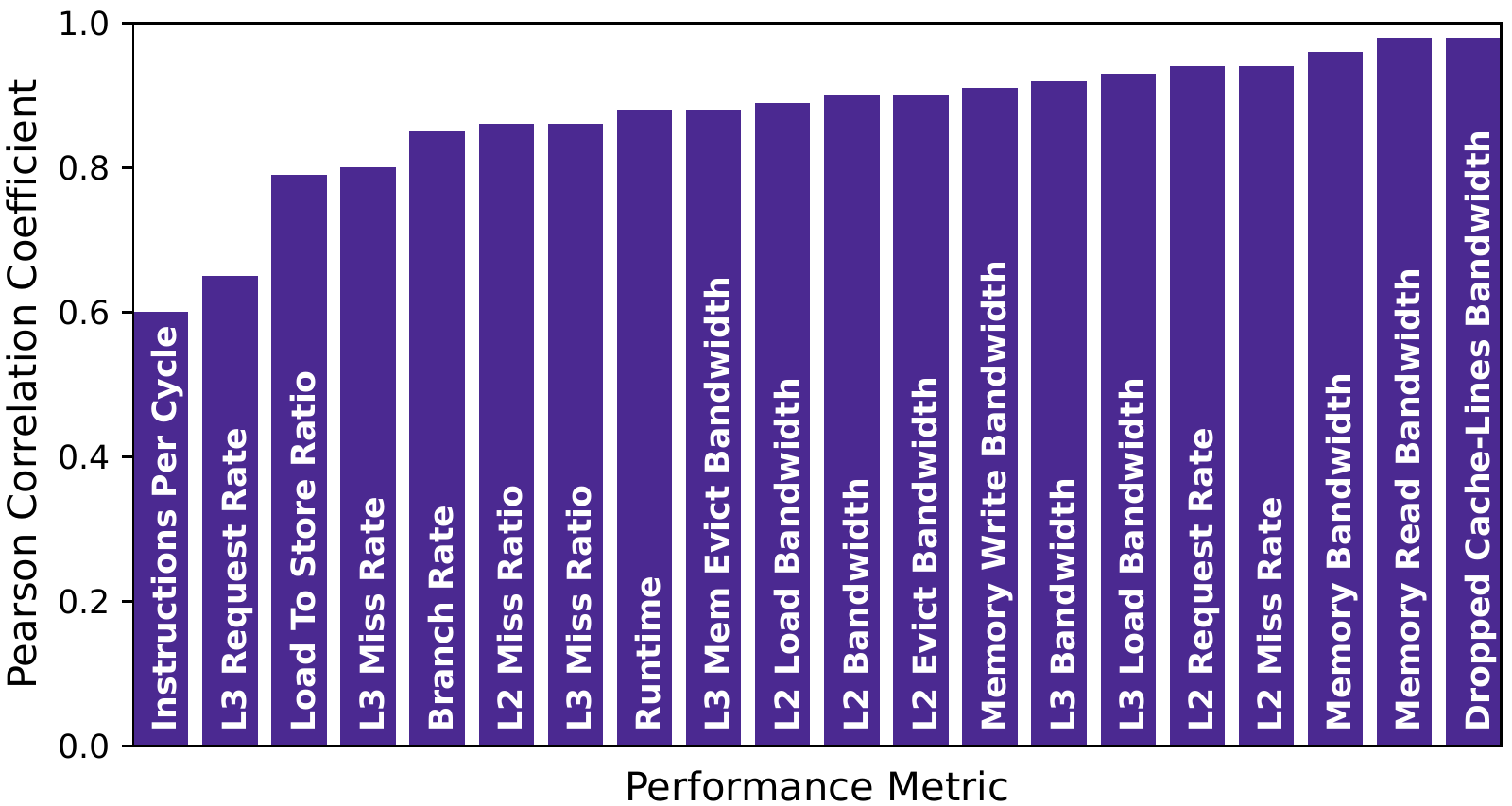}
    \vspace{-2.3em}
    \caption{Pearson correlation coefficient of targets and model predictions.}
    \label{fig:Model_Evaluation}
    \vspace{-1em}
\end{figure}

\paragraph{Targets and Training.}
In order to train the model, we add another linear layer to the model, which predicts a target vector based on the embedding of the parallel loop nest.
These targets comprise $20$ standard performance metrics of the parallel loop nest summarized in Figure~\ref{fig:Model_Evaluation}.
This includes the runtime, the instructions per cycle, different bandwidths, cache miss ratios, and several rates of specific operations per total instructions.
We choose the \textit{mean absolute error} as the loss function and train the model for $20$ epochs using \textit{Adam} at a learning rate of $1\mathrm{e}{-3}$.

\paragraph{Dataset}
We generate the training and validation set synthetically from standard kernels such as maps, reductions, and stencils.
In particular, we include non-data-oblivious kernels such as \textit{boolean masks}.
The test set is extracted from real-world applications implemented in \textit{NPBench}~\cite{Ziogas:2021} by automatically \textit{cutting out} each parallel loop nest.
In total, the sizes of the training, validation, and test sets cover approximately $6$,$500$, $2$,$000$, and $1$,$000$ parallel loop nests, respectively.
In contrast to other models designed to predict the speedup of different schedules, we consider a single \textit{canonical schedule}, which significantly reduces the input variation.
The canonical schedule executes the outermost loop of the loop nest in parallel.

\paragraph{Target Architecture}
The target architecture is an Intel Xeon Gold 6140 CPU with a base clock rate of $2.3$~GHz and $768$ GB of main memory.
The entire dataset is labeled automatically with LIKWID~\cite{Treibig:2014}, which defines groups of performance metrics that can be measured simultaneously.
Each group of metrics is measured in two phases:
In a warmup phase, the program is executed $n_w$ times, where $n_w$ is chosen such that the logical number of bytes moved corresponds to twice the size of the L2 cache but clipped to a maximum of $1$,$000$ repetitions.
In the measurement phase, the program is executed ten times and the median is taken over those measurements to convert the measurements into a single label.
In general, most metrics report the measured mean over all threads.
However, global throughput metrics such as bandwidths or the instruction per cycle are summed over the threads; the runtime is considered as the maximum over all threads.

\subsection{Validation}
Before evaluating the quality of embeddings on application-specific tasks, we validate the model on the prediction of the performance metrics.
Figure~\ref{fig:Model_Evaluation} lists the Pearson correlation coefficient between the targets and the model's predictions on the test set for the different performance metrics.
The minimum correlation of $0.60$ is found for \textit{Instructions Per Cycle} and the maximum correlation of $0.98$ for the metric of \textit{Dropped Cache-Lines Bandwidth}.
For $17$ out of $20$ targets, the correlation is at least $0.80$, indicating a strong correlation between the model prediction and the target labels.

\section{Performance Similarity}
\label{sec:Performance_Similarity}

A similarity search for performance optimization requires that \textit{similar embeddings imply similar performance optimization potentials}.
For instance, if a parallel loop nest has a low memory bandwidth utilization, this loop nest should be mapped to an embedding that is similar to the embeddings of other parallel loop nests with low memory bandwidth utilization.

We evaluate this hypothesis based on the local variation of parallel loop nests under different performance metrics.
Specifically, for each parallel loop nest in the test set, we query the $3$-nearest-neighbors based on the embedding distance and compute the \textit{relative standard deviation} among these four loop nests for a specific performance metric.
We define the mean of the local variations in the test set as the \textit{performance similarity} of the model.

Below, we discuss the similarity metrics we use for our evaluation, the state-of-the-art baselines we compare with, and analyze similarity on the NPBench dataset.

\paragraph{Assessing similarity}
Since the cost for data movement is the dominant factor in performance optimization~\cite{Unat:2016, Unat:2017}, we focus on memory-specific performance metrics for evaluation. The \textit{memory usage efficiency (MUE)}~\cite{Fuhrer:2018} combines the following two performance metrics to assess the optimization potential of a program:
\begin{itemize}
    \item \textit{Main / L3 / L2 Memory Bandwidth}: The attained memory bandwidth on different levels of the memory hierarchy is a standard metric to identify optimization potentials in typical bound-and-bottleneck analyses (cf., \textit{Roofline model}~\cite{Williams:2009, Ilic:2014}). 
    \item \textit{Data Locality}: Fuhrer et al.~\cite{Fuhrer:2018} point out that an analysis based on solely the attained memory bandwidth ignores the intrinsic limitations of the algorithm. For instance, a loop nest with a strided memory access pattern and a loop nest with a random memory access pattern may both yield low memory bandwidths. However, the former may still be optimized through a loop interchange, while the latter already achieves its maximal bandwidth utilization. The data locality accounts for these algorithmic limitations and is defined as the ratio of the \textit{I/O lower bound} \textit{Q} of the algorithm and the measured transferred bytes from main memory \textit{D}, in short, $\frac{Q}{D}$. \textit{Q} is estimated automatically by \textit{SOAP-Analysis}~\cite{Kwasniewski:2021}, which is based on the concept of the \textit{Red-Blue Pebble Game}~\cite{Hong:1981}.
\end{itemize}

\paragraph{Baselines}
In order to assess the model's performance, we compare the similarity of our embeddings with two other models that map parallel loop nests to embeddings.

The \textit{reuse distance analysis}~\cite{Coffman:1973, Beyls:2001, Schaad:2022} is a traditional approach to loop nest analysis, which simulates the execution of the loop for a specified number of iterations on a simplified cache model.
Using this simulation-based analysis, we map each loop nest to a four-dimensional vector of the cache miss ratio, the bytes read from and written to the memory as well as the arithmetic intensity.
The movement of bytes gives a strong indication of the efficiency of the memory access patterns and the arithmetic intensity is typically used to estimate the performance of a program on a target architecture.
Since the simulation of loop nests is expensive, we simulate the first $500$ iterations of the loop nest only.

Baghdadi et. al. \cite{Baghdadi:2021} introduce a state-of-the-art performance model for the optimization of polyhedral programs.
The model estimates the speedup of a schedule and a loop nest based on static features and a \textit{recurrent neural network}.
Since the model is designed to predict the speedup of a certain schedule, we remove the linear prediction layer  and obtain the embedding of the parallel loop nest from the input of this last layer.

\begin{table}[ht!]
    \centering
    \small
        \begin{tabular}{ l r r r r}
            \toprule
            & \multicolumn{3}{c}{Bandwidth} & \multicolumn{1}{c}{Data}\\
            \cmidrule{2-4}
                    & Main & L3 & L2 &\multicolumn{1}{c}{Locality}\\\midrule
            Reuse Distance~\cite{Coffman:1973, Beyls:2001}      & $0.78$    & $1.02$  & $0.82$   & $0.87$\\
            \citet{Baghdadi:2021}   & $0.32$    & $0.41$  & $0.35$   & $0.35$\\
            Our Model               & \textbf{0.25}    & \textbf{0.30}  & \textbf{0.28}   & \textbf{0.31}\\
            \bottomrule
            \hline
        \end{tabular}
  \caption{The mean coefficient-of-variation of different feature extractors on the test set. A lower value means more similarity among the three closest neighbors.}
  \label{table:Homogeneity}
  \vspace{-2em}
\end{table}
\paragraph{Results}
Table~\ref{table:Homogeneity} summarizes the performance similarity of the two baseline feature extractors and our model.
Our model has a strictly lower local variation for all performance metrics and thus yields a higher performance similarity.
Hence, the performance optimization based on the local neighbors in our embedding space is more likely to resolve the actual performance bottlenecks of a parallel loop nest.

\begin{figure}[t]
     \centering
    \begin{subfigure}[b]{.55\linewidth}
        \includegraphics[width=\linewidth]{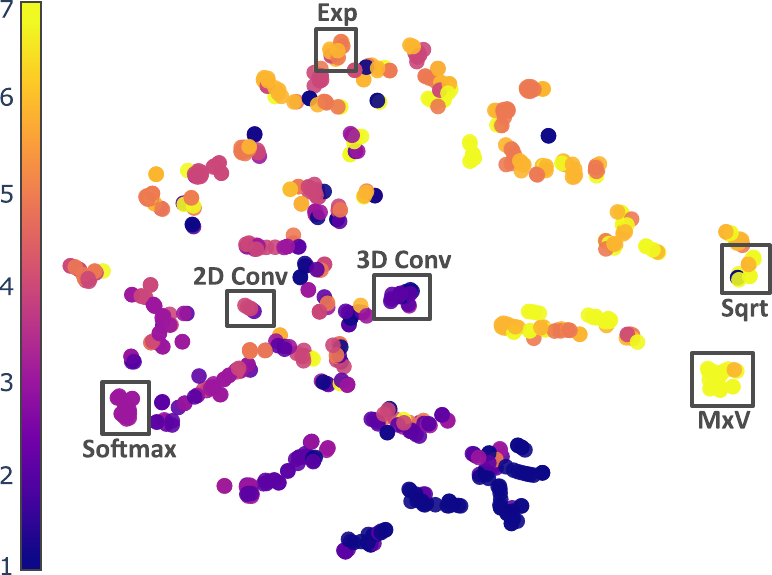}
        \caption{Our Model}
        \label{fig:Homogeneity:perfemb}
    \end{subfigure}
    \hfill
    \begin{subfigure}[b]{.4\linewidth}
        \includegraphics[width=\linewidth]{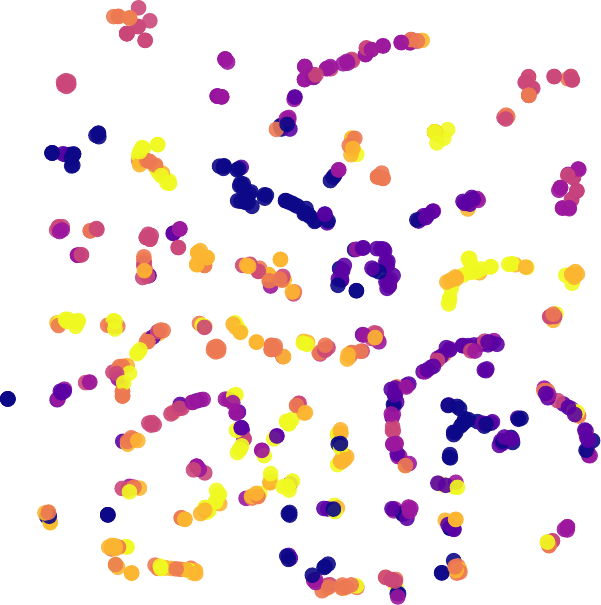}
        \caption{\citet{Baghdadi:2021}}
        \label{fig:Homogeneity:tiramisu}
    \end{subfigure}
    \caption{t-SNE plots of the embedding space generated by our model and \citet{Baghdadi:2021} for the test set. Each sample is colored by the Data Locality MUE metric. The colors are based on binning the range to account for outliers.}
    \label{fig:Homogeneity}
\end{figure}

To further understand the similarity induced by our model, Figure~\ref{fig:Homogeneity} visualizes the embeddings of the test set in a t-SNE plot~\cite{vandermaaten:2008}. A t-SNE plot reduces high-dimensional data onto a 2D plane based on neighborhood minimization. In the figure, each sample is a point colored by its data locality; a plot that is separable by color, as our model's embedding space is (Figure~\ref{fig:Homogeneity:perfemb}), indicates a strong influence of the performance metric in the representation of the sample.
For comparison, Figure~\ref{fig:Homogeneity:tiramisu} shows that the data locality is not an important factor for the representation of the sample, depicted by scattered clusters.

\paragraph{Evaluating importance of static features}
Since the model has a rich set of dynamic features available, the question arises whether the static encoding is actually used by the model.
In order to analyze this question, we analyze the structure of the node embeddings for the input \textbf{array access} nodes of a parallel loop nest.
We extract the node embeddings of input arrays from $350$ synthetically generated parallel loop nests.
For each array, we measure the \textit{L2 load bandwidth} of the \textit{isolated access} to the array.

\begin{figure}[t]
     \centering
    \includegraphics[width=0.67\linewidth]{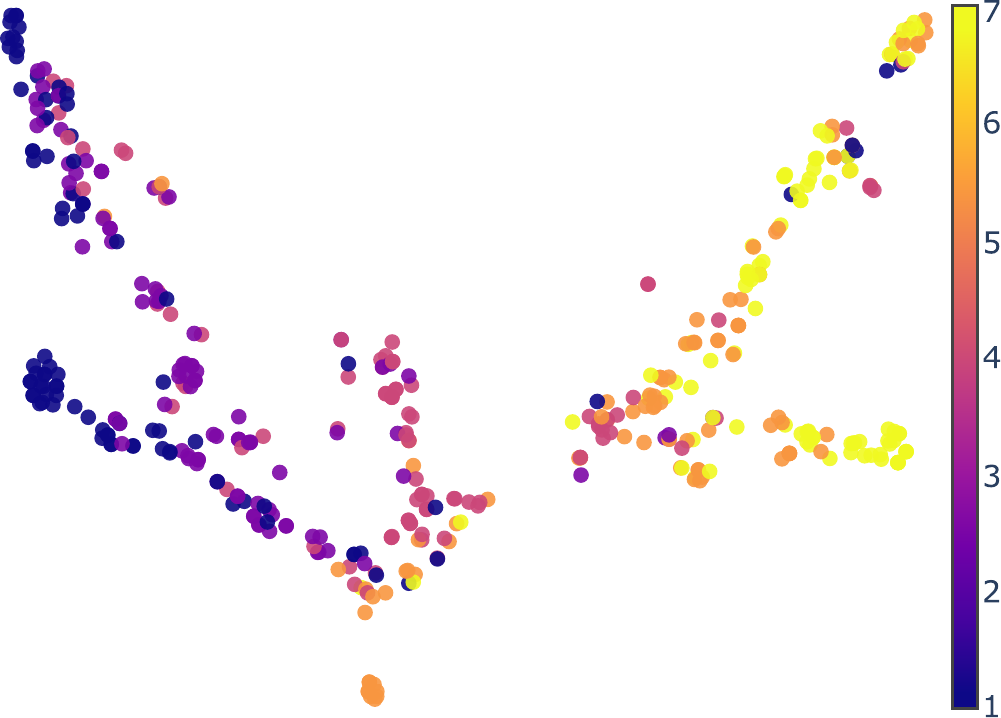}
    \caption{t-SNE plot of the node embeddings of input nodes colored by the L2 load bandwidth. The similarity of local groups indicates that static features of the encoding are utilized by the model. The colors are based on binning the range of the bandwidths to account for outliers.}
    \label{fig:Array_Embeddings}
\end{figure}

We programmatically isolate the access by modifying the parallel loop nests. For example, the isolated access to a matrix $B$ in a matrix-matrix multiplication is shown below:

\begin{center}
    \includegraphics[width=.5\linewidth]{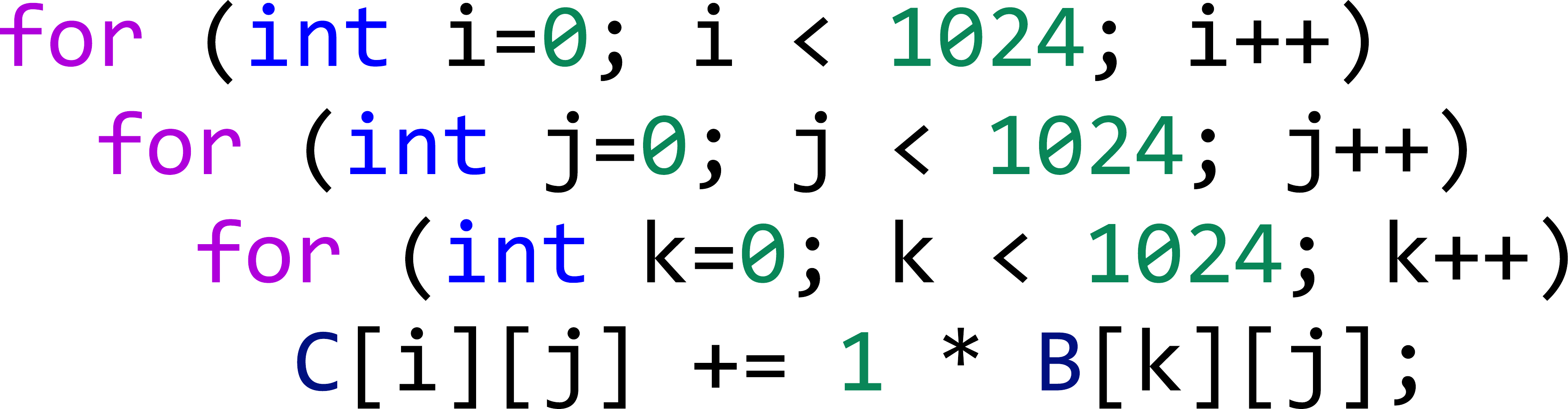}
\end{center}
The resulting t-SNE plot of the node embeddings of input nodes is depicted in Figure~\ref{fig:Array_Embeddings}.
The samples are colored by the measured L2 load bandwidth showing that local groups of node embeddings are similar in the bandwidth of their access.
This indicates that the model generates meaningful embeddings for these nodes based on static features such as the stride of the access and the size of the array.

\section{Transfer Tuning}
\label{sec:transfer_tuning}

Peephole optimization is a compiler technique that operates on a local set of operations.
The replacement rules of peephole optimizations are designed to produce equivalent code and are thus only applicable to a small window of instructions.
Transfer tuning generalizes the idea of peephole optimizations by fuzzy matching program transformations from one subprogram to another via node embeddings.

\subsection{A Matching Problem for Program Transformations}

\begin{figure}[t]
     \centering
    \includegraphics[width=\linewidth]{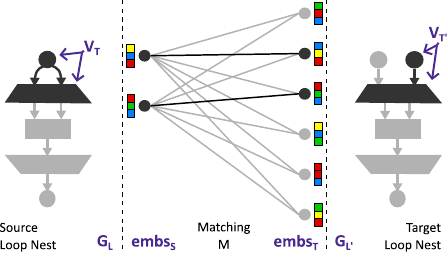}
    \caption{Matching the subgraph of a transformation to another parallel loop nest based on the distances of the node embeddings.}
    \label{fig:Matching}
\end{figure}

Since the representation of the parallel loop nest structure is graph-based, optimizations must be expressed as sequences of transformations on the nodes of a graph.
A transformation can range from a simple change of a node's property to a complex rewrite of a subgraph.
Since the node embeddings generated by the model have a one-to-one correspondence with the nodes in the IR and also have a meaningful structure, each transformation shall be transferred from a source parallel loop nest to a target parallel loop nest by a \textit{matching} of the node embeddings.
In particular, the transfer tuning algorithm (depicted in Figure~\ref{fig:Matching}) consists of four steps:
\begin{enumerate}
    \item Let $G_L = (V_L, E_L)$ be the source parallel loop nest to match and let $G_T = (V_T, E_T)$ be the induced subgraph for a transformation $T$. We compute the source node embeddings $embs_S$ for each $v \in V_T$.
    \item Let $G_{L'} = (V_{L'}, E_{L'})$ be the target parallel loop nest. We compute the target node embeddings $embs_T$ for each $v \in V_{L'}$.
    \item Let $M = (V_{L'},V_T; E, C)$ be a complete, bi-partite graph between the source subgraph's nodes $V_T$ and the target nodes $V_{L'}$, where $C$ is the cost matrix of the pair-wise $\ell_2$ distances between $embs_T$ and $embs_S$. We solve the matching problem $M$ using the \textit{Hungarian method}~\cite{Kuhn:1955} obtaining the mapping between source subgraph's nodes $V_T$ and the target subgraph's nodes $V_T' \subseteq V_{L'}$.
    \item The transformation $T'$ on $G_{L'}$ can now be instantiated from $V_T'$.
\end{enumerate}
For sequences of transformations, the four steps are repeated for every new source and target parallel loop nest of each step.
If the matching problem cannot be solved or the resulting matching does not yield a valid subgraph for the specific transformation, the transformation is skipped.
In practice, we add further constraints to the cost matrix, e.g., setting the cost to infinity for pairs of nodes that do not have the same type.
Furthermore, each transformation requires specific handling of its properties.
For instance, a tiling transformation may not divide the target loop extents evenly.

\section{Evaluation}
\label{sec:Evaluation}

We now evaluate transfer tuning in two case studies:
In the first case study, the optimizations found by a state-of-the-art auto-scheduler for polyhedral applications~\cite{Baghdadi:2021} are transfer tuned between applications from different domains such as image processing, numerical weather prediction, and linear algebra.
In the second case study, dynamic scheduling decisions are transfer tuned between sparse matrix-matrix multiplication (SpMM) for matrices from \textit{suitesparse}~\cite{Davis:2011}.

\subsection{Case Study: Auto-Scheduler}
\label{subsec:Case_Study_I}

Baghdadi et. al.~\cite{Baghdadi:2021} train a speedup prediction model and use this model to guide the search of the Tiramisu auto-scheduler in a large scheduling space consisting of typical loop transformations such as loop interchange, tiling, parallelization, and vectorization.
We show that transfer tuning the discovered optimizations between applications based on the performance embeddings reduces performance optimization to a local search.
The evaluation set consists of $12$ applications comprising approximately one hundred parallel loop nests.

\paragraph{Experimental Setup}
In order to find a strong reference optimization for each parallel loop nest, we run the Tiramisu auto-scheduler's Monte-Carlo Tree Search (MCTS) for a larger number of epochs. Additionally, we test the $100$ best hypotheses found by the search on the target architecture to determine the overall best-performing configuration.
In order to apply this search to our graph IR, we implement a converter from SDFGs to the representation of programs used by the auto-scheduler.
The transfer tuned optimization of each parallel loop nest is found by a $k$-nearest-neighbor search in the embedding space of all parallel loop nests except for the parallel loop nest to be tuned (leave-one-out).

\paragraph{Results}
\begin{table*}
    \centering
        \begin{tabular}{ l r r r r}
            \toprule
            & \multicolumn{2}{c}{\citet{Baghdadi:2021}} & \multicolumn{2}{c}{Transfer Tuning} \\
            \cmidrule{2-3} \cmidrule{4-5}
            & \multicolumn{1}{c}{MCTS Space} & \multicolumn{1}{c}{Runtime [ms]} & \multicolumn{1}{c}{\emph{k=5}} & \multicolumn{1}{c}{\emph{k=10}}\\\midrule
            \textbf{Deep Learning}          & & & &\\
            \emph{mlp}                      & 111,508 & 1.47 & +38.8\% & \textbf{+37.4\%}\\
            \emph{softmax}                  & 183,427 & 110.40 & +0.6\% & +0.5\%\\
            \textbf{Image Processing}       & & & &\\
            \emph{blur filter}              & 1,342 & 1.03 & 0.0\% & 0.0\%\\
            \emph{daubechies wavelet}       & 9,101 & 8.73 & -3.7\% & \textbf{-92.0\%}\\
            \emph{haar wavelet}             & 8,639 & 0.22 & 0.0\% & 0.0\%\\
            \emph{harris filter}            & 1,651 & 9.06 & +0.2\% & -4.0\%\\
            \emph{histogram filter}         & 147,438 & 32.51 & +1.2\% & -4.9\%\\
            \emph{unsharpening filter}      & 25,080 & 29.66 & +3.1\% & +0.5\%\\
            \textbf{Weather Stencils}       & & & &\\
            \emph{heat 3D}                  & 69,080 & 13428.98 & +3.6\% & +2.8\%\\
            \emph{horizontal diffusion}     & 34,534 & 7.00 & +4.8\% & +4.8\%\\
            \textbf{Linear Algebra}         &&&&\\
            \emph{matmul}                   & 65,986 & 14,17 & +5.3\% & +4.1\%\\
            \textbf{Graphs} &&&&\\
            \emph{min-plus mm}              & 65,999 & 24.76 & +9.0\% & +8.5\% \\
            \bottomrule
            \hline
        \end{tabular}
  \caption{The runtime difference of transfer tuning for five and ten neighbors relative to the runtime of the Tiramisu auto-scheduler~\cite{Baghdadi:2021} for polyhedral applications. The auto-scheduler explores a large schedule space using Monte-Carlo Tree Search (MCTS), whereas transfer tuning is a local search based on a few nearest neighbors.}
  \label{table:Evaluation_Transfer_Tuning}
\end{table*}
Table~\ref{table:Evaluation_Transfer_Tuning} lists the results of the Tiramisu auto-scheduler's optimization of each application as well as the results obtained by transfer tuning for $k=5$ and $k=10$ neighbors.
For the majority of applications, the transfer-tuned runtime is within $5\%$ of the reference at a fraction of the search complexity, see \textit{MCTS Space} column for the number of configurations tested by the auto-scheduler.
Since the reference optimizations are found once and then stored in the database, transfer tuning enables exhaustive offline optimization of applications with a large scheduling space.

\paragraph{Daubechies Wavelet}
In the embedding space, the neighbors of a parallel loop nest act as a collection of explored search paths based on slightly varied input conditions.
The \textit{Daubechies wavelet} benchmark is an example where this neighborhood yields a considerable speedup.
The application consists of a single parallel loop nest, where the outermost loop iterates over the $3$ channels of an image.
Parallelizing over this loop induces a major performance bottleneck on a CPU with $36$ cores, since most of the cores are idling.

Upon inspecting the transferred transfer tuning results, we see that it optimized according to the \textit{Haar wavelet}:
MCTS fails to find an optimization maximizing the parallelism for the Daubechies wavelet, but succeeds to find an optimization for the almost identical Haar wavelet. This also showcases an important feature of performance embeddings --- as opposed to end-to-end neural networks, \textbf{\textit{transfer tuning provides explainability for its optimization decisions}}.

Other examples are the \textit{Harris filter} and the \textit{histogram filter}, in which transfer tuning finds additional potential for applying the optimization found within the same benchmark.

\paragraph{Multi-Layer Perceptron (MLP)}
Although \textit{matmul} and \textit{min-plus matrix multiplication} are potential candidates for optimizing the layers in \textit{mlp}, we see that transfer tuning performs worse for this particular benchmark. 
The matrix multiplications of matmul and mlp differ significantly in the dimensions of the matrices:
while \textit{matmul} multiplies a $1024\times2048$ and a $2048\times1024$ matrix, mlp multiplies weight matrices, which have a small leading dimension of $64$ corresponding to the batch size.
Hence, the matrix multiplications define different trade-offs of data locality and parallelization.
This shows that the density (i.e., the availability of similar neighbors) of the optimization database for certain applications is an important hyperparameter of transfer tuning.

\subsection{Case Study: Tailored Optimization}

In the second case study, we demonstrate the extensibility of transfer tuning to custom optimizations on the example of dynamically scheduling SpMMs for matrices from suitesparse~\cite{Davis:2011}.
A typical performance bottleneck of SpMM is an imbalanced distribution of work among the threads, resulting from the distribution of the non-zero elements.
The standard optimization is then to change the scheduling from a \textit{static} assignment of work to threads to a \textit{dynamic} assignment, which incurs some overhead for the execution.

\paragraph{Experimental Setup}
In order to define an optimization database for the scheduling decision, we determine the optimal schedule for $42$ sparse matrices from suitesparse~\cite{Davis:2011} by benchmarking \textit{OpenMP's} default static schedule and a dynamic schedule of chunk size $8$.
The matrices are multiplied by a dense matrix of $512$ columns filled with random values.
We evaluate whether transfer tuning can decide the optimal schedule by splitting this set of matrices into a set that is stored in the optimization database and a test set.
The scheduling of the test set matrices is then determined by a $1$-nearest neighbor query to the database.
The resulting runtime of the matrices is compared with the \textit{Intel MKL 2021.3} implementation of SpMM.

\paragraph{Results}
\begin{figure}[ht!]
    \centering
    \includegraphics[width=0.4\textwidth]{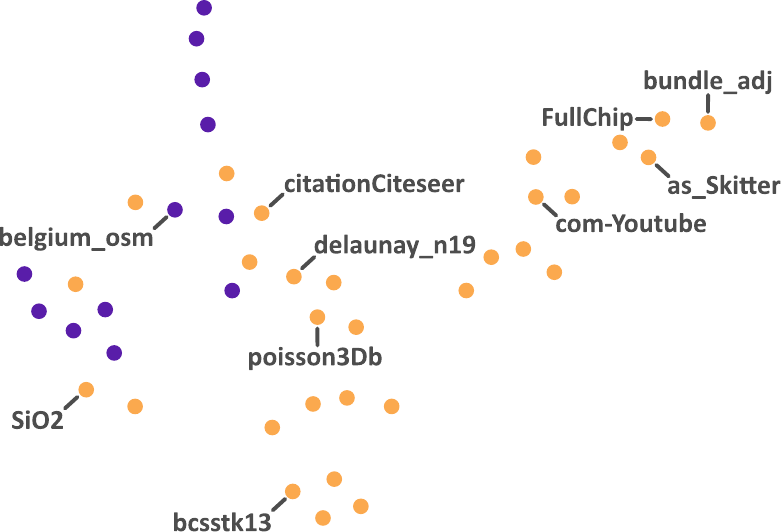}
    \caption{t-SNE plot of the SpMM embeddings for $42$ suitesparse matrices. Embeddings are colored by the optimal scheduling type, i.e., static (purple \colordot[magmapurple,fill=magmapurple]{.3em}) and dynamic (orange \colordot[magmaorange,fill=magmaorange]{.3em}).}
    \label{fig:Suitesparse_scheduling}
\end{figure}
The t-SNE plot of the SpMM embeddings of all matrices is depicted in Figure~\ref{fig:Suitesparse_scheduling}, where the embeddings of the different matrices are colored by their optimal schedule.
The separation of groups by colors already indicates the applicability of the $1$-nearest-neighbor approach to dynamic scheduling.
Table~\ref{table:Suitesparse} summarizes the runtimes of both schedules, the runtime after transfer tuning as well as the Intel MKL baseline.
Transfer tuning picks the correct scheduling decision for $8$ out of the $10$ test benchmarks.
Furthermore, the comparison with the Intel MKL baseline shows a significant speedup of the optimal scheduling for a different subset of $8$ out of $10$ benchmarks.
\begin{table}
\small
    \centering
        \begin{tabular}{ l r r | r r}
            \toprule
            Sparse Matrix & Static & Dynamic & Transfer & MKL \\\midrule
            \emph{as-Skitter}               & 2574.19 & 719.31 & \textbf{719.31} & 1264.84\\
            \emph{delaunay\_n19}            & 132.46  & 111.78 & 111.78 & \textbf{101.59}\\
            \emph{poisson3Db}               & 157.94 & 86.00 & \textbf{86.00} & 112.79\\
            \emph{citationCiteseer}         & 135.59 & \textbf{125.43} & 135.59 & 134.23\\
            \emph{FullChip}                 & 4081.38 & 3028.05 & \textbf{3028.05} & 3863.76\\
            \emph{belgium\_osm}             & 180.88 & 206.83 & \textbf{180.88} & 240.03\\
            \emph{com-YouTube}              & 911.14 & 286.34 & \textbf{286.34} & 392.93\\
            \emph{bcsstk13}                 & 2.90 & 1.82 & 1.82 & \textbf{0.62}\\
            \emph{bundle\_adj}              & 4395.73 & 437.39 & \textbf{437.39} & 840.43\\
            \emph{SiO2}                     & 450.68 & \textbf{174.84} & 450.68 & 263.09\\
            \bottomrule
            \hline
        \end{tabular}
  \caption{Runtime of SpMM for the static and the dynamic scheduling in the left part of the table and the runtime of transfer tuning and Intel MKL in the right part of the table.
  }
  \label{table:Suitesparse}
\end{table}

\begin{figure}[t]
    \centering
    \includegraphics[width=\linewidth]{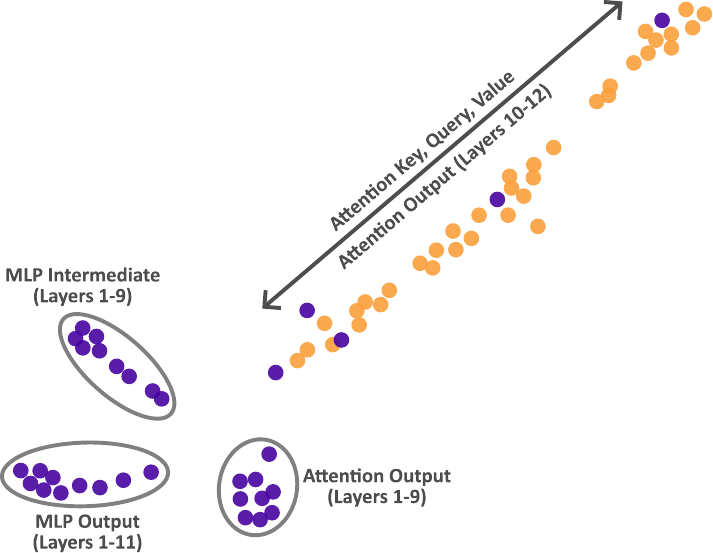}
    \caption{t-SNE plot of the SpMM embeddings for the sparse weights of a BERT model~\cite{Kurtic:2022}. The embeddings are colored by the optimal scheduling type, i.e., static (purple \colordot[magmapurple,fill=magmapurple]{.3em}) and dynamic (orange \colordot[magmaorange,fill=magmaorange]{.3em}).}
    \label{fig:BERT_scheduling}
    \vspace{-.5em}
\end{figure}

\paragraph{BERT}
The BERT transformer~\cite{bert} is a standard neural network architecture in natural language processing.
The sparsification of the dense layers is a common technique to enable efficient inference by sacrificing a reasonable amount of accuracy~\cite{Hoefler:2022}.
In order to show the cross-domain transfer of this knowledge, we repeat the above experiment for the sparse weights of a sparsified model~\cite{Kurtic:2022}, yielding a similarly separable embedding space for transfer tuning.
The tSNE plot of the sparse weights is depicted in Figure~\ref{fig:BERT_scheduling}.

In conclusion, transfer tuning yields comparable performance speedups on all tested cases, at times outperforming existing tools and libraries by inferring cross-application optimizations. It can adapt to additional insights gained by automated tools and tailored optimizations and can be inspected to explain its reasoning behind certain optimizations via the chosen neighbor.

\section{Related Work}

Automatic performance optimization and performance modeling for optimization has been studied by a variety of works.
The following section summarizes prior related research.

\paragraph{Performance Modeling and Extrapolation}
Several wo-rks focused on the automatic prediction of program and subprogram performance.
One of the earlier instances of using machine learning for performance modeling was performed by \citet{ipek-ann-predict}, who use an MLP to predict application performance.
\citet{carrington-predict} and \citet{siegmund-performance-influence} also provide performance prediction for tuning via heuristic means on an application-level, and 
\citet{calotoiu-modeling} model and extrapolate runtime dependency on parameters of general codes via time measurement of multiple small experiments. Most such works do not focus on the optimization transformations and their choice, but rather on accurate execution time prediction.

Application-specific performance models~\cite{modesto,wu-trace-extrapolation,li-stencil} introduce domain knowledge into the prediction and often use the generated communication or performance model to inform an optimization search without executing the program, which might be expensive due to running on distributed environments.

\paragraph{Polyhedral Compilers}
 The \textit{Pluto}~\cite{Bondhugula:2008}, \textit{PENCIL}~\cite{Baghdadi:2015}, and \textit{LLVM Polly}~\cite{Grosser:2012} compilers express performance optimizations as the solution of an integer linear program (ILP) with respect to a hand-crafted cost model of the target architecture.
For reasons of tractability of the ILP, the cost model makes strong simplifying assumptions, often yielding sub-optimal results on complex architectures~\cite{Baghdadi:2013}.

\paragraph{Deep Code Representations}
\textit{inst2vec}~\cite{BenNun:2018}, \citet{Brauckmann:2020}, and \textit{ProGraML}~\cite{Cummins:2021} are examples of neural code representations that map static code to embeddings.
The embeddings are designed to solve typical compiler tasks and classify applications according to their \textit{semantics}.
In contrast, performance embeddings encode both static and dynamic properties, with the express goal of capturing performance aspects regardless of the underlying algorithm.

\paragraph{Optimizing Compilers}
Optimizing compilers are subject to extensive research.
\textit{Tiramisu}~\cite{Baghdadi:2019}, \textit{Halide}~\cite{Ragan-Kelley:2018} and \textit{TVM}~\cite{Chen:2018} introduce deep learning performance models~\cite{Adams:2019, Baghdadi:2021, Chen:2018} based on static features, which guide the search in the scheduling space within the subset of supported programs.
\citet{Singh:2022} extends these performance models to graph neural networks improving the accuracy of the prediction.
\citet{Steiner:2021} re-formulate the search problem as a \textit{Markov Decision Problem}, which can be solved using reinforcement learning.
Other works utilize input-specific and profiling features to automatically optimize sparse linear algebra routines~\cite{Elafrou:2017}.
Our approach separates the performance model from the optimization by introducing an offline optimization database.
This allows the local search in the application space, which significantly reduces the complexity of the search and allows for the extension of the optimization space without re-training the model.

\paragraph{Transfer Tuning}
\citet{Martins:2016} cluster C functions based on static features to select the optimal compiler passes according to the cluster assignment.
\citet{Gibson:2022} provide a constrained definition of the term transfer tuning as the reuse of optimizations found by auto-schedulers for specific operations in tensor programs.
The discovered optimizations are matched by hand-crafted heuristics to other operations.
Our approach extends this concept to intermediate representations and optimizations based on a fuzzy matching of node embeddings.
The similarity of performance embeddings thereby generalizes hand-crafted transfer rules. 

\section{Discussion}
\label{sec:Discussion}

The following section briefly discusses possible extensions of the presented similarity-based framework.

\paragraph{Scalability}
The density of the optimization database is a crucial hyperparameter for the validity of the similarity-based approach.
However, the separation of the model and the transformations enables offline search for further optimizations.
This allows to continuously improve the quality of the search by extending the database (i.e., \textit{online learning}) with suboptimal examples.
For existing auto-schedulers, a corresponding extension of the approach means expensive re-training and a significant increase in the scheduling space for all applications.
This is a practical problem since current auto-schedulers often fail for basic applications, such as the \textit{jacobi2d} benchmark on the model of \citet{Baghdadi:2021} or the \textit{max filter} on \citet{Adams:2019}.
A possible next step for the approach is to evaluate transfer tuning with larger databases.

\paragraph{Transformation Alignment}
The matching algorithm matches a transformation to a parallel loop nest using the Hungarian method.
However, the matching of a sequence of transformations is modeled greedily, which means that a database is required that covers symmetric cases as separate entries.
However, such cases typically require a simple modification of the transformation sequence.
For instance, a loop interchange, which is a common infix in transformation sequences, may often be skipped or replaced by a similar interchange for specific pairs of loop nests.
This problem could be modeled as a \textit{sequence alignment problem}, where the skipping or insertion of specific transformations are latent decisions (represented by, e.g., a \textit{Hidden Markov Model}).
Sequence alignments are well-known in the field of \textit{machine translation}~\cite{Och:1999, Vaswani:2017}.
Understanding performance optimization as a sequence alignment between a reference optimization and a similar loop nest gives rise to the idea of a model-based alternative to the model-free reinforcement learning approach presented by \citet{Steiner:2021}. 

\paragraph{Loop Fusion}
The fusion of parallel loop nests is an important optimization to reduce the volume of necessary data movement.
In order to support this optimization in the similarity-based framework, a model is necessary which produces \textit{subgraph embeddings} for graphs of parallel loop nests.
Such models are subject to current research~\cite{Alsentzer:2020}.

\paragraph{Target Architecture}
The separation of the model and the optimizations also facilitates porting the approach to new architectures.
In particular, learning a representation for similarity search is significantly simpler than training a model that accurately predicts the speedups of complex optimization sequences.
In fact, the dynamic encoding and the targets only need to be substituted by appropriate performance counters and metrics for the new target architecture.
Performance models usually provide a good basis for finding relevant metrics and are available for most architectures, e.g., NUMA nodes~\cite{Denoyelle:2018}, FPGA~\cite{daSilva:2013}, GPU~\cite{Nugteren:2012}, and distributed computing~\cite{Culler:1993}.

\section{Conclusion}

In this paper, we present a similarity-based tuning framework that lifts peephole optimizations by fuzzy-matching larger program transformations.
The approach separates the performance model from the optimizations in the form of performance embeddings and an optimization database.
This enables local search for optimizations over the nearest neighbors in the embedding space.

We demonstrate the approach in different case studies highlighting the reduction of the search complexity by up to four orders of magnitude, and the extensibility of the approach to tailored optimizations on data-dependent applications, outperforming the state-of-the-art MKL library in certain use cases.
The approach creates a new space that can be used for explainable and robust optimization, while remaining adaptive to future applications and hardware --- transferring a new optimization technique is as simple as adding a row to the database.

\begin{acks}
This work received EuroHPC-JU funding with support from the European Union’s Horizon 2020 program and from the European Research Council under grant agreement PSAP, number 101002047.
The authors also wish to acknowledge the support from the PASC program (Platform for Advanced Scientific Computing) for the DaCeMI project.
T.B.N. and P.S. were supported by the Swiss National Science Foundation (Ambizione Project \#185778).
\end{acks}

\bibliographystyle{ACM-Reference-Format}
\bibliography{literature}
\clearpage

\appendix

\section{Appendix}
\label{sec:Appendix}

The static encoding maps nodes and edges of an SDFG to a set of features.
The mapping of SDFG node types to features is summarized in Table~\ref{table:Static_Features}.
\begin{table}[!ht]
     \centering
     \resizebox{\columnwidth}{!}{%
     \begin{tabular}{l p{6cm}}
     \toprule
     \emph{Node Type} & \emph{Features} \\
     \midrule
     Access Node    & data type, bytes per element, shape, total size, stride, alignment, offset, transient, storage type\\
     \hline
     Map Entry      & map level, map dimensions, map extents, map steps\\
     \hline
     Map Exit       & \textit{one-hot encoding}\\
     \hline
     Memlet         & start access matrix, stop access matrix, steps vector, dynamic, indirection, reduction, type of reduction\\
     \bottomrule
     \hline
     \end{tabular}
     }
     \caption{An overview of the static features selected for the static encoding of parallel loop nests. Most features directly correspond to the properties of nodes in an SDFG.}
    \label{table:Static_Features}
\end{table}

The dynamic encoding maps the profiling to $19$ performance counters selected from $8$ different groups.
Table~\ref{table:Performance_Counters} lists the counters and groups in detail.
\begin{table}[!ht]
    \centering
    \resizebox{\columnwidth}{!}{%
    \begin{tabular}{l l}
    \toprule
    \emph{Group} & \emph{Counters} \\ \midrule
    Instructions & INSTR\_RETIRED\_ANY\\
     \hline
    \multirow{4}{*}{FP 32}
     & FP\_ARITH\_INST\_RETIRED\_SCALAR\_SINGLE\\
     & FP\_ARITH\_INST\_RETIRED\_128B\_PACKED\_SINGLE\\
     & FP\_ARITH\_INST\_RETIRED\_256B\_PACKED\_SINGLE\\
     & FP\_ARITH\_INST\_RETIRED\_512B\_PACKED\_SINGLE\\
     \hline
    \multirow{4}{*}{FP 64}
     & FP\_ARITH\_INST\_RETIRED\_SCALAR\_DOUBLE\\
     & FP\_ARITH\_INST\_RETIRED\_128B\_PACKED\_DOUBLE\\
     & FP\_ARITH\_INST\_RETIRED\_256B\_PACKED\_DOUBLE\\
     & FP\_ARITH\_INST\_RETIRED\_512B\_PACKED\_DOUBLE\\
     \hline
    \multirow{2}{*}{Branching}
     & BR\_INST\_RETIRED\_ALL\_BRANCHES\\
     & BR\_MISP\_RETIRED\_ALL\_BRANCHES\\
     \hline
    \multirow{2}{*}{DRAM Controller}
     & MEM\_INST\_RETIRED\_ALL\_LOADS\\
     & MEM\_INST\_RETIRED\_ALL\_STORES\\
     \hline
    \multirow{2}{*}{Main Memory}
     & CAS\_COUNT\_RD\\
     & CAS\_COUNT\_WR\\
     \hline
    \multirow{2}{*}{L3 Cache}
     & L2\_LINES\_IN\_ALL\\
     & L2\_TRANS\_L2\_WB\\
     \hline
    \multirow{2}{*}{L2 Cache}
     & L1D\_REPLACEMENT\\
     & L1D\_M\_EVICT\\
     \bottomrule
     \hline
    \end{tabular}
    }
     \caption{An overview of the performance counters selected for the dynamic encoding on the Intel Xeon Gold 6140 CPU.}
    \label{table:Performance_Counters}
\end{table}

\end{document}